\documentclass[%
 aip,
 pop,
 amsmath,amssymb,
 reprint,%
 floatfix
]{revtex4-2}

\usepackage{graphicx}
\usepackage{dcolumn}
\usepackage{bm}
\usepackage[hidelinks]{hyperref}
\usepackage{physics}
\usepackage{esint}
\usepackage[utf8]{inputenc}
\usepackage[T1]{fontenc}
\usepackage{mathptmx}
\usepackage{etoolbox}

\makeatletter
\def\@email#1#2{%
 \endgroup
 \patchcmd{\titleblock@produce}
  {\frontmatter@RRAPformat}
  {\frontmatter@RRAPformat{\produce@RRAP{*#1\href{mailto:#2}{#2}}}\frontmatter@RRAPformat}
  {}{}
}%

\makeatother

\newcommand{\vv}[1]{{\boldsymbol{#1}}}

\providecommand{\uv}[1]{\ensuremath{\boldsymbol{\hat{#1}}}}

\begin{document}

\title{
    Spherical compression of an applied magnetic field in inertial confinement fusion
}

\author{R. Spiers}
\author{A. Bose}
\email{bose@udel.edu}
\author{C. A. Frank}
\affiliation{Department of Physics and Astronomy, University
of Delaware, Newark Delaware 19716, USA}
\author{D. J. Strozzi}
\author{J. D. Moody}
\author{C. A. Walsh}
\author{B. A. Hammel}
\affiliation{Lawrence Livermore National Laboratory, 7000 East Ave, Livermore California 94550, USA}

\date{\today}

\begin{abstract}
Applying an external magnetic field to laser-driven inertial confinement fusion implosions is a promising approach for enhancing fusion yield. The field is compressed with the plasma, producing a magnetized hotspot that anisotropically suppresses thermal losses and traps alpha particles, making performance sensitive to the compressed field orientation.
We derive a simple, readily applicable analytic model that enables rapid evaluation of the compressed field topology and show that ablation into the hotspot amplifies the central field, while the ablated ice near the hotspot edge develops a decaying, radially bent field, with a discontinuity in the field direction. The radially bent field renders thermal insulation at the hotspot edge negligible and largely independent of the applied field strength, whereas insulation in the hotspot core still depends strongly on the applied field. 
Applying the model to non-axial initial field configurations, we find that an initially applied mirror field provides the greatest suppression, followed by the standard axial field.
\end{abstract}

\maketitle

\section{Introduction} \label{sec:intro}

One concept to enhance the performance and repeatability of inertial confinement fusion (ICF) implosions \cite{atzeniPhysicsInertialFusion2009,lindlInertialConfinementFusion1998,nuckollsLaserCompressionMatter1972} is to apply an external magnetic field ($\sim$1-50 T) which insulates against thermal losses \cite{braginskiiTransportProcessesPlasma1965,perkinsPotentialImposedMagnetic2017,walshMagnetizedICFImplosions2022,spiersHotspotModelInertial2025}, enhances alpha-particle confinement in the hotspot \cite{guskovEnergyTransferParticles1984a,knappEffectsMagnetizationFusion2015}, and is hypothesized to suppress mixing of degradatory ablator material into the fuel \cite{srinivasanMitigatingHydrodynamicMix2014,walshMagnetizedAblativeRayleighTaylor2022,walshResistiveDiffusionMagnetized2024a}. In magnetized ICF, the seed field is approximately ``frozen-in'' to the ionized plasma and compressed along with the implosion to $\sim 10^3$x its original strength, magnetizing the central high-temperature ``hotspot''.

Presently, there are only two disparate methods for predicting the magnetic field structure in magnetized ICF implosions---complex radiation-magnetohydrodynamics (MHD) simulations and a simple scaling relation. In the former, computationally-intensive 2D/3D multiphysics simulations are performed for specific ICF designs, producing complex field topologies where the effect of each physics term is difficult to isolate. This includes simulations of room-temperature \cite{strozziDesignModelingIndirectly2024} and layered \cite{perkinsPotentialImposedMagnetic2017,djordjevicIntegratedRadiationMagnetoHydrodynamicSimulations2026a} implosions in the 2D MHD code \texttt{Lasnex}, and 2D/3D \texttt{Gorgon} simulations providing scaling studies \cite{walshMagnetizedICFImplosions2022}, non-axial field geometries \cite{walshMagnetizedICFImplosions2025a}, design optimization \cite{walshMagnetizedICFImplosions2025b}, burn-wave propagation \cite{oneillBurnPropagationMagnetized2025a}, and instability calculations \cite{walshPerturbationModificationsPremagnetisation2019a}. On the other side of the spectrum, the average magnetic field in the hotspot can be predicted using the capsule convergence ratio. Considering a loop around the equator of the capsule, Faraday's law with ideal Ohm's law ($\vv{E} = -\vv{u} \times \vv{B}$) implies that magnetic flux is conserved \cite{gotchevMagnetoinertialApproachDirectdrive2008a} and so nominally $R_0^2 B_{0,z} \approx R_f^2 B_{f,z}$, where $R_0$ and $R_f$ are the initial and final waist radii and $B_{0,z}$ and $B_{f,z}$ are the axial magnetic field strengths in the equatorial plane ($z=0$). For typical ICF convergence ratio of 30, this simple calculation predicts ($B_{f,z}=(R_f/R_0)^2 B_{0,z}$) a 900x enhancement in the field strength during compression. This oversimplified model, however, assumes that an initial axial (cylindrical) $B_{0,z}$ field continues to be perfectly axial $B_{f,z}$ and homogeneous after a (spherical) compression of the capsule.

In this work, we bridge the gap between oversimplified scaling relations and computationally-intensive MHD simulations by providing an analytic model for the leading-order contribution to the compressed field. We expect this to be useful for a number of reasons: (1) fundamental physics of magnetized implosions, (2) rapid analysis of magnetized ICF designs, (3) usefulness in semi-analytic hotspot models \cite{hurricaneInertiallyConfinedFusion2016,springer3DDynamicModel2018,caseyMultirocketPistonModel2025,spiersHotspotModelInertial2025}, (4) proper initial conditions for magnetized burn studies, and (5) code validation for new radiation-MHD codes. Point (4) is especially important, since there is disagreement in the literature about whether magnetization suppresses the propagation of fusion burn waves and therefore suppresses yield. Using symmetrized capsule simulations, O'Neill \textit{et al.} \cite{oneillBurnPropagationMagnetized2025a} finds that magnetizing the burn suppresses yield by 42\% with respect to no applied field using the \texttt{Gorgon} code, however integrated hohlraum simulations in Djordjevi\'c \textit{et al.} \cite{djordjevicIntegratedRadiationMagnetoHydrodynamicSimulations2026a} show that magnetized burn increases yield in each studied capsule design. Other modeling efforts initialize the simulations with a prescribed magnetic field at peak compression, like Jones and Mead \cite{jonesPhysicsBurnMagnetized1986} with homogeneous azimuthal fields, Sefkow \textit{et al.} \cite{sefkowDirectlyDrivenMagnetized2024} creating an \textit{ad hoc} magnetic field profile for compressed fast-ignition targets, or Xu \textit{et al.} \cite{xuControlBurningWave2025} who use an axial magnetic field in the hotspot. As we will show in this work, these prescribed field profiles are not realistic for an imploded capsule. Since magnetized burn is sensitive to the angle between the magnetic field and the burn propagation direction, it is important that future studies of burn in magnetized designs use a physically-motivated compressed magnetic field to improve the reliability of the results.

In this work, we describe an exact analytic solution to the full spatially-resolved magnetic field under spherical compression or expansion. This model, henceforth denoted as \textit{advection-only}, gives the influence of the fluid on the magnetic field without the feedback of the field on the fluid evolution (e.g., through magnetic tension or thermal conduction). The advection-only model is rapid, flexible, and will explain the complex phenomena observed in the field profiles generated by 2D radiation-MHD ICF simulations. 

This paper is laid out as follows: in Sec. \ref{sec:theory} we derive the conservation laws for the magnetic field under spherical flows, resulting in an exact solution for the compressed magnetic field geometry. Then in Sec. \ref{sec:results}, we show that mass ablation from the dense shell into the hotspot strengthens the central field while weakening and bending the field in the ablated ice. The effects of additional MHD terms are discussed. We apply the model to non-symmetric implosions with low-mode asymmetries (mode-2, mode-4) in Sec. \ref{sec:nonspherical}. Finally, in Sec. \ref{sec:nonaxial}, the model is used to predict the compressed field structure of various non-axial initial field topologies, suggesting that a mirror field might provide moderately better thermal insulation in the core of the hotspot. A user-code is described in the Appendix \ref{subsec:usercode} and has been made publicly available on \texttt{GitHub} \cite{spiersBoseHEDPField_compression2026}.

\section{Field advection} \label{sec:theory}

The evolution of a magnetic field in plasmas is governed by the induction equation. The equations of extended-MHD are constructed using a generalized Ohm's law \cite{braginskiiTransportProcessesPlasma1965,sadlerSymmetricSetTransport2021}, but many of these terms drop out when the implosion is axially symmetric (e.g., cross-gradient Nernst, Righi-Leduc). We also remove terms that only produce localized fields around instabilities in ICF (e.g., Biermann \cite{walshSelfGeneratedMagneticFields2017,frankSelfgeneratedMagneticFields2024,farmerSimulationSelfgeneratedMagnetic2017a} and Z-gradient source terms \cite{sadlerMagnetizationMixJets2020}). The remaining terms affecting magnetic field evolution include advection, resistive diffusion, and Nernst advection.
\begin{multline} \label{eq:induction}
    \frac{\partial \vv{B}}{\partial t} = 
    \nabla \times \left( \vv{u} \times \vv{B} \right) 
    - \nabla \times \left( D_\eta \nabla \times \vv{B} \right) \\
    - \nabla \times \left( \gamma_\perp \frac{\tau}{m_e} \nabla T_e \times \vv{B} \right)
\end{multline}
In this equation, $\vv{u}$ is the fluid velocity, $D_\eta = \left( 25.95 \mathrm{\ cm^2 \ keV^{3/2} \ s^{-1}} \right) \alpha_\parallel \bar{Z} \log \Lambda / T^{3/2}$ is the resistive diffusion coefficient with $\bar{Z}$ the average ionization, $\alpha_\parallel = 0.506$ for $\bar{Z}=1$, and $\log \Lambda$ is the Coulomb logarithm. The relevant transport coefficients $\alpha_\parallel$ and $\gamma_\perp$ are tabulated \cite{sadlerSymmetricSetTransport2021} as a function of $\bar{Z}$ and the electron Hall parameter $\chi \equiv \omega_{ce} \tau$, where $\omega_{ce}$ is the electron cyclotron frequency and $\tau$ is the electron-ion collision time.

In this analytic derivation, we consider only the advection of the field due to bulk fluid flow (the $\nabla \times (\vv{u} \times \vv{B})$ term), and discuss corrections for the remaining smaller terms in Sec. \ref{sec:results}. Since $\nabla \cdot \vv{B} = 0$, the ideal MHD induction equation can be expanded as 
\begin{equation} \label{eq:advectiveinduction}
    \frac{\partial \vv{B}}{\partial t} + \vv{u} \cdot \nabla \vv{B} + \vv{B} \nabla \cdot \vv{u} - \vv{B} \cdot \nabla \vv{u} = 0.
\end{equation}
These terms can be evaluated by assuming that the flows are radial (i.e., $\vv{u} = u_r(r,\theta,\varphi)\uv{r}$), which permits angular dependence of the radial velocity, for example, if the pole converges faster than the equator. The components of the last directional derivative in Eq. \ref{eq:advectiveinduction} can then be expressed in spherical coordinates as 
\begin{multline}
    \vv{B} \cdot \nabla \vv{u} = 
    B_r \left( \nabla \cdot \vv{u} \right) \uv{r} 
    + \left( \frac{B_\theta}{r} \frac{\partial u_r}{\partial \theta} + \frac{B_\varphi}{r \sin\theta} \frac{\partial u_r}{\partial \varphi} \right) \uv{r} \\
    + \frac{u_r}{r} \left( -2 B_r \uv{r} + B_\theta \uv{\theta} + B_\varphi \uv{\varphi} \right)
\end{multline}
Using the material derivative ($\frac{D}{Dt} \equiv \frac{\partial}{\partial t} + \vv{u} \cdot \nabla$), the radial component of Eq. \ref{eq:advectiveinduction} becomes 
\begin{equation}
    \frac{D B_r}{Dt} + \frac{2u_r}{r} B_r = \frac{B_\theta}{r} \frac{\partial u_r}{\partial \theta} + \frac{B_\varphi}{r \sin\theta} \frac{\partial u_r}{\partial \varphi},
\end{equation}
where the right-hand-side is only non-zero when the flows are asymmetric---for a spherically symmetric implosion $u_r$ is independent of $\theta$ and $\varphi$. Using the property that $u_r \equiv \frac{D r}{Dt}$ (since $\vv{u} \cdot \nabla r = u_r$) and integrating by parts, the general (i.e. $u_r(r,\theta,\varphi)$) conservation law for $B_r$ can be written as 
\begin{equation} \label{eq:conservationBr}
    \frac{1}{r^2} \frac{D}{Dt} \left( r^2 B_r \right) = \frac{B_\theta}{r} \frac{\partial u_r}{\partial \theta} + \frac{B_\varphi}{r \sin\theta} \frac{\partial u_r}{\partial \varphi},
\end{equation}
demonstrating that $r^2 B_r$ is conserved in ideal spherically compressing or expanding flows (i.e. $u_r(r)$). This result is well-known for spherical flows \cite{parkerDynamicsInterplanetaryGas1958,weberAngularMomentumSolar1967}, as one can consider flux conservation through a hemisphere or differential solid angle whose area is proportional to $r^2$.

The $\theta$ and $\varphi$ components of Eq. \ref{eq:advectiveinduction} are equivalent to one another, yielding 
\begin{equation} \label{eq:advectiveBtheta}
    \frac{D B_{\theta/\varphi}}{Dt} + B_{\theta/\varphi} \nabla \cdot \vv{u} - B_{\theta/\varphi} \frac{u_r}{r} = 0.
\end{equation}
The velocity divergence is simplified using the mass continuity equation, namely that
\begin{equation} \label{eq:continuitylagrangian}
    \rho \frac{D}{Dt} \left( \frac{1}{\rho} \right) = \nabla \cdot \vv{u},
\end{equation}
which, upon insertion to Eq. \ref{eq:advectiveBtheta} and integration by parts yields the conservation laws for $B_\theta$ and $B_\varphi$.
\begin{equation} \label{eq:conservationBtheta}
    \frac{D}{Dt} \left( \frac{B_{\theta/\varphi}}{\rho r} \right) = 0
\end{equation}

Given an arbitrary 3D initial magnetic field profile, one can use the conservation laws in Eqs. \ref{eq:conservationBr} and Eqs. \ref{eq:conservationBtheta} to determine the final \textit{advection-only} field profile under compression/expansion irrespective of the path. For illustrative purposes, we assume the compression is spherical ($\frac{\partial u_r}{\partial \theta} = \frac{\partial u_r}{\partial \varphi} = 0$) although in Sec. \ref{sec:nonspherical} we demonstrate the alternative solution including asymmetry corrections. Denoting $R_0$ as the initial position and $R_f$ as the final position of the fluid element, then the field is given by 
\begin{equation} \label{eq:evolutionBsph}
    \vv{B}_f = \left( \left( \frac{R_0}{R_f} \right)^2 \uv{r} \uv{r} + \left( \frac{\rho_f R_f}{\rho_0 R_0} \right) \left( \uv{\theta} \uv{\theta} + \uv{\varphi} \uv{\varphi} \right) \right) \cdot \vv{B}_0,
\end{equation}
where $\vv{B}_0$ and $\rho_0$ are evaluated at $R_0$ and $\vv{B}_f$, $\rho_f$ correspond to the new fluid positions at $R_f$. These fluid locations can be found using a Lagrangian radiation-hydrodynamics simulation or, equivalently, by using the initial and final density profiles, recognizing that the mass of a Lagrangian fluid element $M_\mathrm{enc}$ is constant in time.
\begin{equation} \label{eq:lagrangian}
    M_\mathrm{enc} = \int_0^{R_0} \rho_0 x^2 dx = \int_0^{R_f} \rho_f x^2 dx
\end{equation}
The evolution equation (\ref{eq:evolutionBsph}) can also be derived using the vector potential representation of the induction equation. This derivation is demonstrated in the Appendix (\ref{subsec:vecpot}) but is less general since it only applies to symmetric spherical compression ($u_r(r)$) and with $B_\varphi = 0$.

It is convenient to recast Eq. \ref{eq:evolutionBsph} into cylindrical coordinates ($s$-$z$, so as to not confuse the spherical radius with the cylindrical radius), as in magnetized ICF an axial $\vv{B}_0 = B_0 \uv{z}$ field is often applied. In doing so, one inserts $\uv{r} = \sin \theta \uv{s} + \cos \theta \uv{z}$ and $\uv{\theta} = \cos \theta \uv{s} - \sin \theta \uv{z}$ into Eq. \ref{eq:evolutionBsph}. We also drop $B_\varphi$ in this expression since its evolution is exactly the same as in spherical coordinates.
\begin{equation} \label{eq:evolutionBcyl}
\begin{pmatrix}
    B_{s f} \\
    B_{z f}
\end{pmatrix} = \left( \frac{R_0}{R_f} \right)^2 
\begin{pmatrix}
    1 - \alpha \cos^2 \theta  & \alpha \sin\theta \cos\theta \\
    \alpha \sin\theta \cos\theta  & 1 - \alpha \sin^2 \theta
\end{pmatrix}
\begin{pmatrix}
    B_{s 0} \\
    B_{z 0}
\end{pmatrix}
\end{equation}
Here, we have defined the parameter $\alpha$ as 
\begin{equation} \label{eq:alpha}
    \alpha \equiv 1 - \frac{\rho_f R_f^3}{\rho_0 R_0^3} 
\end{equation}
In this representation, the compression of the field depends on two fundamental parameters: the per-cell convergence ratio $R_0/R_f$ and the parameter $\alpha$ which measures how much a Lagrangian cell has stretched in the radial direction. From Eq. \ref{eq:alpha} it is apparent that the evolution of $\alpha$ is given by 
\begin{equation} \label{eq:alphaconservation}
    \frac{D}{Dt} \left( \frac{1-\alpha}{\rho R^3} \right) = 0,
\end{equation}
where $\alpha(t=0) = 0$. After some manipulation and making use of Eq. \ref{eq:continuitylagrangian}, one can show that $\alpha$ evolves according to 
\begin{equation} \label{eq:alphaEOM}
    \frac{D \alpha}{Dt} = (1-\alpha) r \frac{\partial}{\partial r} \left( \frac{u_r}{r} \right).
\end{equation}
In homogeneous compression, where each fluid element converges at the same rate, based on the self-similar Sedov-Taylor solution, the velocity profile is linear\cite{kidderTheoryHomogeneousIsentropic1974,atzeniPhysicsInertialFusion2009} (i.e. $u_r \propto r$). Therefore, Eq. \ref{eq:alphaEOM} shows that $\alpha$ remains 0 in homogeneous compression and the field lines do not bend (see Eq. \ref{eq:evolutionBcyl}); this is also evident from Eq. \ref{eq:alphaconservation} which shows that for a homogeneous compression $\rho R^3$ is conserved, and thus $\alpha = 0$. However, under ablating flows with material blow-off, the velocity profiles rise slower than linearly since mass conservation at the ablation surface requires $\rho_1 u_1 = \rho_2 u_2$, implying that the implosion velocity inside the hotspot is greater than immediately inside the shell. This non-linearity in the hotspot velocity profile causes $\alpha$ to asymptotically increase towards 1, upon which $B_{\theta f}$ and $B_{\varphi f}$ become small compared to the radial component of the field $B_{r f}$ (see Eq. \ref{eq:evolutionBsph}).

As an example, consider the compression of an axial magnetic field, $\vv{B}_0 = B_0 \uv{z}$. Using Eq. \ref{eq:evolutionBcyl}, the compressed field is
\begin{equation} \label{eq:axialfieldstructure}
    \vv{B}_f = B_0 \left( \frac{R_0}{R_f} \right)^2 \left( 
        \alpha \sin \theta \cos \theta \uv{s} + \left( 1 - \alpha \sin^2 \theta \right) \uv{z} \right),
\end{equation}
indicating that the field remains axial ($\vv{B} \cdot \uv{s} = 0$) unless $\alpha$ is non-zero. The magnitude of this field is given by 
\begin{equation} \label{eq:axialfieldmagnitude}
    \left| \vv{B}_f \right| = B_0 \left( \frac{R_0}{R_f} \right)^2 \sqrt{1 - \alpha \left( 2 - \alpha \right) \sin^2 \theta},
\end{equation}
which implies that the magnitude of the compressed axial field depends on the polar angle only if $\alpha \neq 0$. Furthermore, in the ablated regions where $\alpha \to 1$, this implies that 
\begin{equation} \label{eq:axialfieldmagnitude_lineout}
    \left| \vv{B}_f \right| \to B_0 \left( \frac{R_0}{R_f} \right)^2 \cos \theta,
\end{equation}
meaning that the field in this region will be strongest at the pole and decrease to zero near the equator.

\section{Results} \label{sec:results}

In Sec. \ref{sec:theory}, we derived the equations for the field profile as it is advected with radially compressing flows. This general field compression model is also applicable to non-axial initial fields and for departures from spherical compression ($u_r(r, \theta, \varphi)$). In this section we demonstrate the resulting field profile for the simplest case: axial starting field, spherical compression, and no feedback of the magnetic field on the implosion dynamics. We show that the structure of the field inside the hotspot is radically altered when there is ablation from the shell into the hotspot (as occurs in DT-ice layered capsules) versus in implosions with no ablation into the hotspot (as in NIF Symcaps \cite{moodyIncreasedIonTemperature2022a,sioPerformanceScalingApplied2023,strozziDesignModelingIndirectly2024} or exploding pushers \cite{boseEffectStronglyMagnetized2022a}).

\begin{figure}[ht] 
    \centering 
    \includegraphics[width=3.4in]{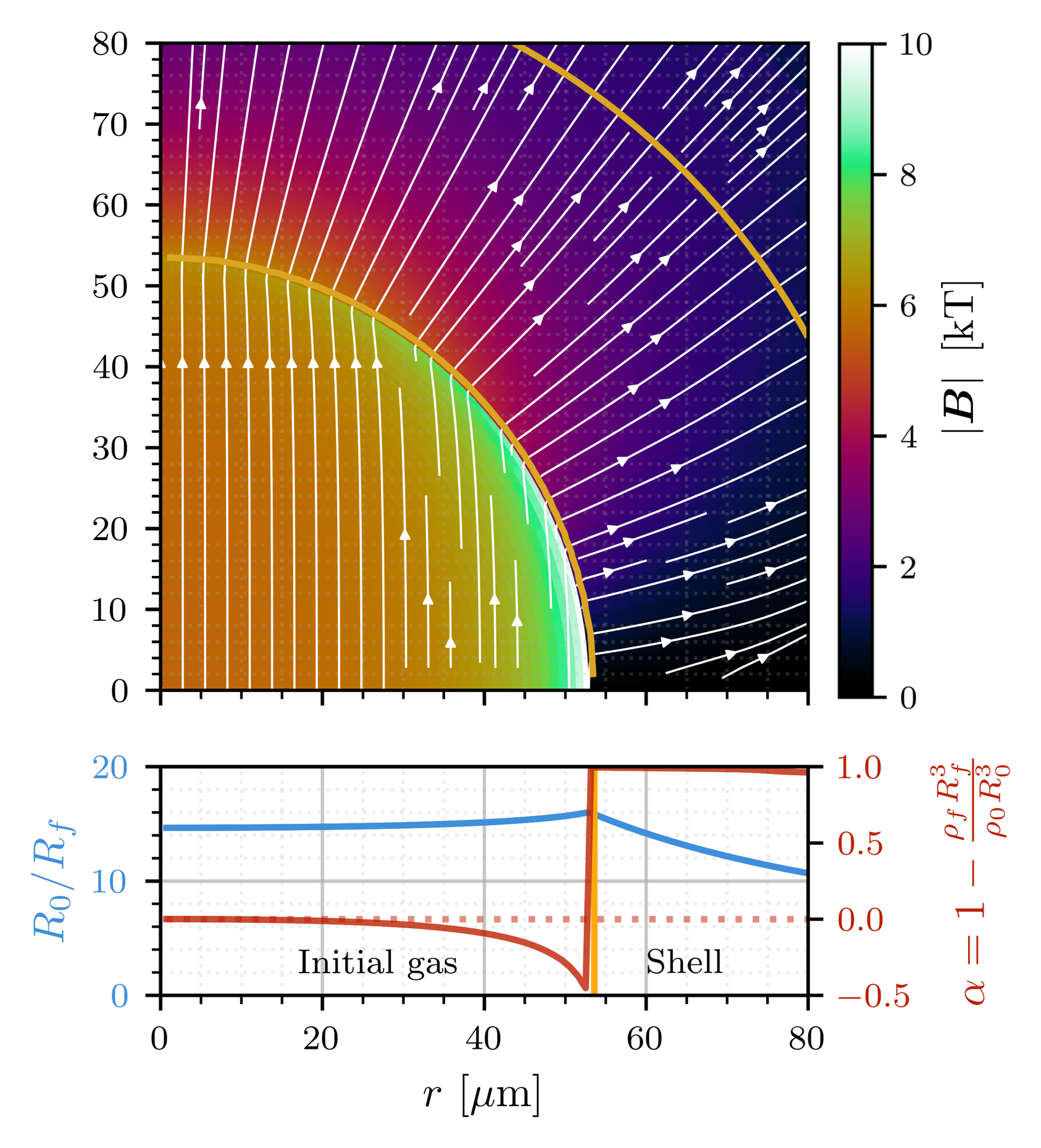}
    \caption{(Top) Magnetic field profile at peak compression (7.5 ns) resulting from applying the advection-only model (Eq. \ref{eq:evolutionBcyl}) to NIF Symcap N210607 with an axial applied field. The inner and outer shell boundaries are marked with orange lines. Field lines are shown in white, with background heatmap showing the field magnitude in kT. (Bottom) Field model variables $R_0/R_f$ (convergence ratio, blue) and $\alpha$ (bending parameter, red) used in the model, determined using a 1D HYDRA simulation. The vertical orange line represents the shell inner boundary.}
    \label{fig:fieldbt_N210607} 
\end{figure}

Fig. \ref{fig:fieldbt_N210607} shows the resulting field profile and field compression variables ($R_0/R_f$ and $\alpha$) for the NIF Symcap N210607 which has no ablation from the HDC shell into the D$_2$ hotspot. Since the compression is dominantly homogeneous, the hotspot fluid elements converge the same amount ($R_0/R_f \approx 17$) and the field-lines are not substantially bent. However, inside the shell, thermal expansion and outer-surface ablation have bent the field-lines in the radial direction ($\alpha$ asymptotically approaching 1). This feature of magnetized implosions without ablation into the hotspot---that hotspot fields remain axial and are amplified by a constant $(R_0/R_f)^2$ factor---is important to justify the common assumption \cite{spiersHotspotModelInertial2025,walshMagnetizedICFImplosions2022,sioPerformanceScalingApplied2023} for NIF Symcaps that $\vv{B}_f = B_0 \left( R_0 / R_f \right)^2 \uv{z}$. However, as will be shown in the following text, this does not hold for DT-ice layered implosions.

\begin{figure}[ht] 
    \centering 
    \includegraphics{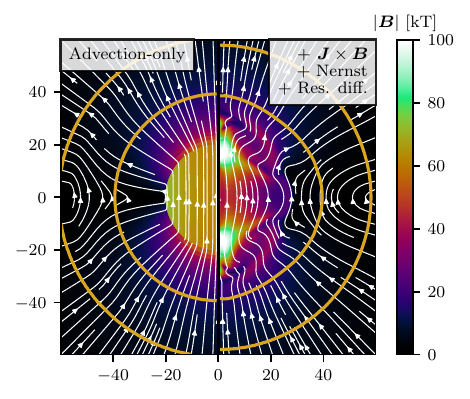} 
    \caption{Magnetic field profiles at peak compression (9.11 ns) for modeling and simulation of NIF DT-layered shot N210808 with a 26 T applied axial field. The colormap corresponds to the field magnitude (in kT) and streamlines show the field direction. Axis units are in $\mathrm{\mu m}$. Yellow contours show the position of the shell as a contour of the density $\rho_\mathrm{max}/e$. Results correspond to (left) the advection-only analytic model from Eq. \ref{eq:evolutionBcyl}, and (right) with feedback on the implosion through the $\vv{J} \times \vv{B}$ force and field evolution from Nernst advection and resistive diffusion. Neither panel includes anisotropic thermal conduction or alpha-particle deposition.} 
    \label{fig:fieldbt_N210808} 
\end{figure}

The remainder of the results in this work are focused on DT-ice layered NIF experiment N210808 which first exceeded the Lawson criterion \cite{zylstraExperimentalAchievementSignatures2022b}, although the conclusions drawn on the magnetic field compression if an external field were to be applied are generalizable to most hotspot-ignition NIF capsule designs. We perform capsule-only simulations \cite{weberMixingICFImplosions2020} of the N210808 design which are driven using a symmetrized X-ray frequency-dependent-source from integrated hohlraum simulations. Almost all results are generated by post-processing 1D \texttt{HYDRA} \cite{marinakThreedimensionalSimulationsNova1996} Lagrangian radiation-hydrodynamics simulations using Eq. \ref{eq:evolutionBcyl}, although Fig. \ref{fig:fieldbt_N210808} provides direct results from 2D extended-magnetohydrodynamics (XMHD) simulations. Currently \texttt{HYDRA} lacks the capability for XMHD on a spherical polar mesh, so these 2D simulations are run with no magnetic field until 7.3 ns (after shock breakout but before shock convergence), where the magnetic field is initialized using Eq. \ref{eq:evolutionBcyl} on a cylindrical butterfly mesh \cite{clarkModelingAblatorDefects2024}, then run through peak compression (9.11 ns). Each simulation is performed assuming the initial applied field strength is 26 T, consistent with the magnetized ICF experiments fielded on the NIF \cite{moodyIncreasedIonTemperature2022a,sioPerformanceScalingApplied2023}.

\begin{figure}[ht] 
    \centering 
    \includegraphics{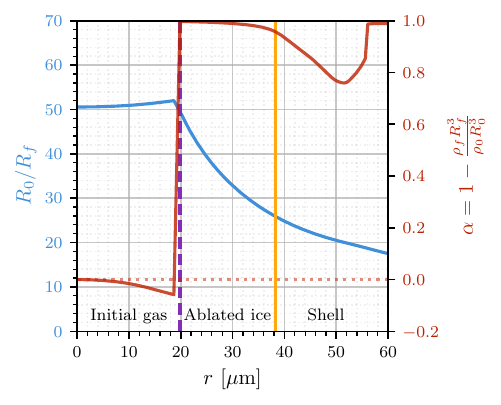} 
    \caption{Convergence ratio ($R_0/R_f$, left) and field bending parameter ($\alpha$, right) at peak compression from a 1D simulation of DT-layered shot N210808. The vertical lines correspond to the gas/ablated ice interface (purple dashed line) and the hotspot/shell interface (orange solid line).}
    \label{fig:alphaCR} 
\end{figure}

The field compression parameters $\alpha$ and $R_0/R_f$ are demonstrated in Fig. \ref{fig:alphaCR}, corresponding to the hotspot during peak compression of N210808. During the deceleration phase, thermal conduction from the hotspot into the shell causes mass ablation into the hotspot \cite{bettiHotspotDynamicsDecelerationphase2001}. Even though the commonly-quoted convergence ratio ($R_\mathrm{gas}/R_\mathrm{hs}$) is 26, the zone corresponding to the initial gas fill converges by 50x, creating much stronger fields in the central core. However, the zone corresponding to the ablated ice has $\alpha \to 1$, causing the field to become purely radial (see Eq. \ref{eq:evolutionBsph}) and rapidly decrease in magnitude. Additionally, there is a sharp discontinuity in $\alpha$ at this interface between initial gas and ablated ice, originating from the initial density discontinuity in the capsule before implosion. 

Starting with an initial 26 T axial applied field, we can compute the resulting compressed field profiles using Eq. \ref{eq:evolutionBcyl}. Shown in Fig. \ref{fig:fieldbt_N210808}, this causes a region of high magnetic field in the initial gas region ($\sim$80 kT out to 20 $\mathrm{\mu m}$) that remains axial, and a rapidly decaying predominantly radial field ($\vv{B} \sim r^{-2} \cos \theta \uv{r}$) in the ablated ice, as it was derived in Eq. \ref{eq:axialfieldmagnitude_lineout}. Since the magnetic field is in the radial direction in the outer hotspot and in the shell, this has substantial consequences on the degree of thermal insulation, alpha-heating, and magnetized burn suppression. As was pointed out in simulations by O'Neill \textit{et al.},\cite{oneillBurnPropagationMagnetized2025a} magnetic fields which are perpendicular to the shell do not meaningfully suppress the propagation of burn waves into the cold fuel shell.

We now turn our attention to the interface in Fig. \ref{fig:fieldbt_N210808} between the initial gas region and the ablated ice which occurs at $r = 20 \ \mathrm{\mu m}$. At this interface, $\alpha$ transitions from 0 to 1, which according to Eqs. \ref{eq:axialfieldstructure} and \ref{eq:axialfieldmagnitude}, causes a discontinuity in both the magnetic field direction ($\vv{B} \propto \uv{z}$ to $\vv{B} \propto \uv{r}$) and field strength. Magnetic tension ($\frac{1}{\mu_0}\vv{B} \cdot \nabla \vv{B}$) resists this rapid change in field direction, since the small radius of curvature of the field lines leads to large magnetic tension forces. 

Whereas the left panel of Fig. \ref{fig:fieldbt_N210808} shows the field profile only from advection of the field with the implosion, the right panel includes feedback of the field on the evolution of the plasma via the Lorentz force ($\vv{J} \times \vv{B}$), Nernst advection, and resistive diffusion. This field structure from the 2D extended-MHD simulation differs from the analytic solution in that magnetic energy is concentrated at the poles and weaker along the waist, field-lines are curved around the implosion core, and localized strong bending occurs in the ablated ice. We claim that these differences result from the gas/ablated ice discontinuity, which has jump conditions consistent with a hybrid \textit{rotational discontinuity} and \textit{magnetosonic wave}. The rotational discontinuity arises from the discontinuity in field direction, and propagates as a large-amplitude Alfv\'en wave via magnetic tension throughout the hotspot at the local Alfv\'en speed ($v_A = |B|/\sqrt{\mu_0 \rho}$), which is approximately $50-100 \ \mathrm{\mu m/ns}$ (moving $\approx 10 \ \mathrm{\mu m}$ during the deceleration phase). In contrast, the magnetosonic wave moves much faster at the local sound speed ($\approx 500 \ \mathrm{\mu m/ns}$) and, analogous to how thermal pressure becomes isobaric during subsonic hotspot formation, the total pressure ($p + B^2/2 \mu_0$) adjusts via the magnetosonic waves to become homogeneous throughout the hotspot, although the flows and corresponding influence on the magnetic field topology is small. The propagation dynamics of the rotational discontinuities in a magnetized ICF hot-spot are complex and will be described at length in a separate follow-up companion paper to this work. In this paper, we focus on the leading order effect, which is the advection of the field lines and the effect of mass ablation at the inner shell interface causing a radial bending of the field lines at the hot-spot edge (i.e. initial gas and ablated-ice interface).

The relevance of other extended-MHD terms can also be estimated for this setup. Magnetic pressure is a small effect for implosions without large initial fields ($B_0 \lesssim 50 \ \mathrm{T}$) \cite{perkinsPotentialImposedMagnetic2017}, quantified by the plasma $\beta = p/(B^2/2\mu_0) \approx 251 \left( p \ [\mathrm{Gbar}] \right) / \left( B \ [\mathrm{kT}] \right)^2$, where $\beta \gg 1$ implies that fluid pressure dominates over magnetic pressure. Based on the results in Fig. \ref{fig:fieldbt_N210808}, $\beta$ in the initial gas region is 10--100 throughout the implosion and is $\sim$10x larger in the ablated ice (where the field is weaker). At peak compression, $\beta \approx 20$ in the initial gas region, implying that magnetic pressure is not a dominant effect but may be an important correction term. The Nernst velocity is small ($u_N < 1 \ \mathrm{\mu m/ns}$, compared to implosion velocity of 100's of $\mathrm{\mu m/ns}$ ) in the hotspot since the electron Hall parameter is large ($\chi > 10$, causing $\gamma_\perp < 0.01$). Finally, the hotspot magnetic diffusivity is $D_\eta \sim 1 \ \mathrm{\mu m^2 / ns}$ (magnetic Reynolds number $\mathrm{Re_m} > 5000$) implying the field is frozen-in to the bulk hotspot (diffusion time $\approx 100$ ns) but diffusion may impact small-scale fields or decay fields in the cold, dense shell (where $\mathrm{Re_m} < 200$ until the onset of the deceleration phase).

\section{Non-spherical compression} \label{sec:nonspherical}

Another feature of the presented advection-only model is that it can be used to determine the field structure due to asymmetric compression of the capsule. Low-mode shape asymmetries are a common degradation source in NIF \cite{caseyDiagnosingOriginImpact2023e,sioPerformanceScalingApplied2023} and OMEGA \cite{boseEffectStronglyMagnetized2022a} implosions caused by asymmetries in the x-ray or laser drive. These asymmetries, described by Legendre polynomials $P_n (\cos \theta)$, grow exponentially during the deceleration phase \cite{caseyDiagnosingOriginImpact2023e}. Taking the shell deceleration $g$ to be approximately constant, the evolution of the shell for a single-mode ($n$) asymmetry can be written as 
\begin{equation} \label{eq:shellasymmetry}
    R \left( \theta, t \right) = R_\mathrm{1D} (t) \left( 1 + e^{ \sqrt{\frac{g}{R_\mathrm{1D}}} \left( t - t_\mathrm{pc} \right) } p_n P_n(\cos \theta) \right),
\end{equation}
where $p_n$ is the Legendre coefficient of the stagnated shell at time of peak compression $t_\mathrm{pc}$, and $R_\mathrm{1D}$ is the symmetric radius. In Bose \textit{et al.} it was shown that the non-radial velocity contribution to residual kinetic energy was smaller than the radial residual kinetic energy ($< 20\%$ of total) for low-mode, low-amplitude asymmetries \cite{bosePhysicsLongIntermediatewavelength2017b}. Under the limiting assumption that compression remains radial (i.e., $u_\theta = 0$, but allowing asymmetric $u_r(r, \theta, \varphi)$), then Eq. \ref{eq:conservationBr} can be numerically integrated to determine the resulting field profiles.

\begin{figure}[ht] 
    \centering 
    \includegraphics{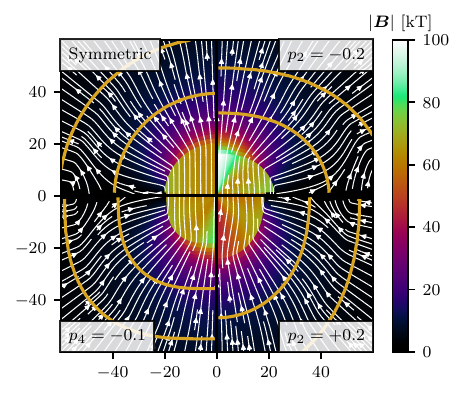} 
    \caption{Magnetic field profiles for N210808-like asymmetrically-compressed implosions, found by numerically integrating Eqs. \ref{eq:conservationBr} and \ref{eq:conservationBtheta} with prescribed asymmetry given by Eq. \ref{eq:shellasymmetry}.} 
    \label{fig:asymmetric} 
\end{figure}

The magnetic field profiles for common low-mode asymmetry configurations is shown in Fig. \ref{fig:asymmetric}. In all scenarios, the shape of the region of high field, corresponding to the initial gas, is the same as the surrounding distorted shell, although the field magnitude is redistributed. For the oblate case ($p_2 < 0$), the field along the pole becomes stronger since it is more compressed; the opposite is true for the prolate case $p_2 > 0$. Although the maximum field increases substantially in the oblate case, the average field within the hotspot is nearly unchanged.

It is known that magnetized anisotropic thermal conduction causes the hotspot to become more prolate by suppressing heat transport at the waist \cite{walshMagnetizedICFImplosions2022,strozziDesignModelingIndirectly2024,oneillBurnPropagationMagnetized2025a,walshMagnetizedICFImplosions2025b}. This intrinsic feature of magnetization will further influence the magnetic field profiles near the ablation surface, although this is a higher-order effect that is not accounted for in our simple advection-only model, and will require a future integrated field compression model.

\section{Non-axial fields} \label{sec:nonaxial}

One option when designing magnetized ICF ignition implosions is the structure of the applied magnetic field. Although the magnetized ICF experiments have all used axial fields generated by a Helmholtz-like coil, it has been suggested that non-axial fields may produce more optimal thermal insulation and alpha-trapping characteristics \cite{walshMagnetizedICFImplosions2025a}. In Walsh \textit{et al.} \cite{walshMagnetizedICFImplosions2025a}, they found using 2D XMHD simulations that applying a ``mirror'' field to NIF layered implosion N170601 caused the hotspot temperature to increase by 60\% compared to the $B=0$ case, whereas an axial field only enhances the temperature by 40\%. These simulations did not include alpha-heating, which may further enhance the performance of the mirror field.

\begin{figure*}[ht] 
    \centering 
    \includegraphics[width=6.0in]{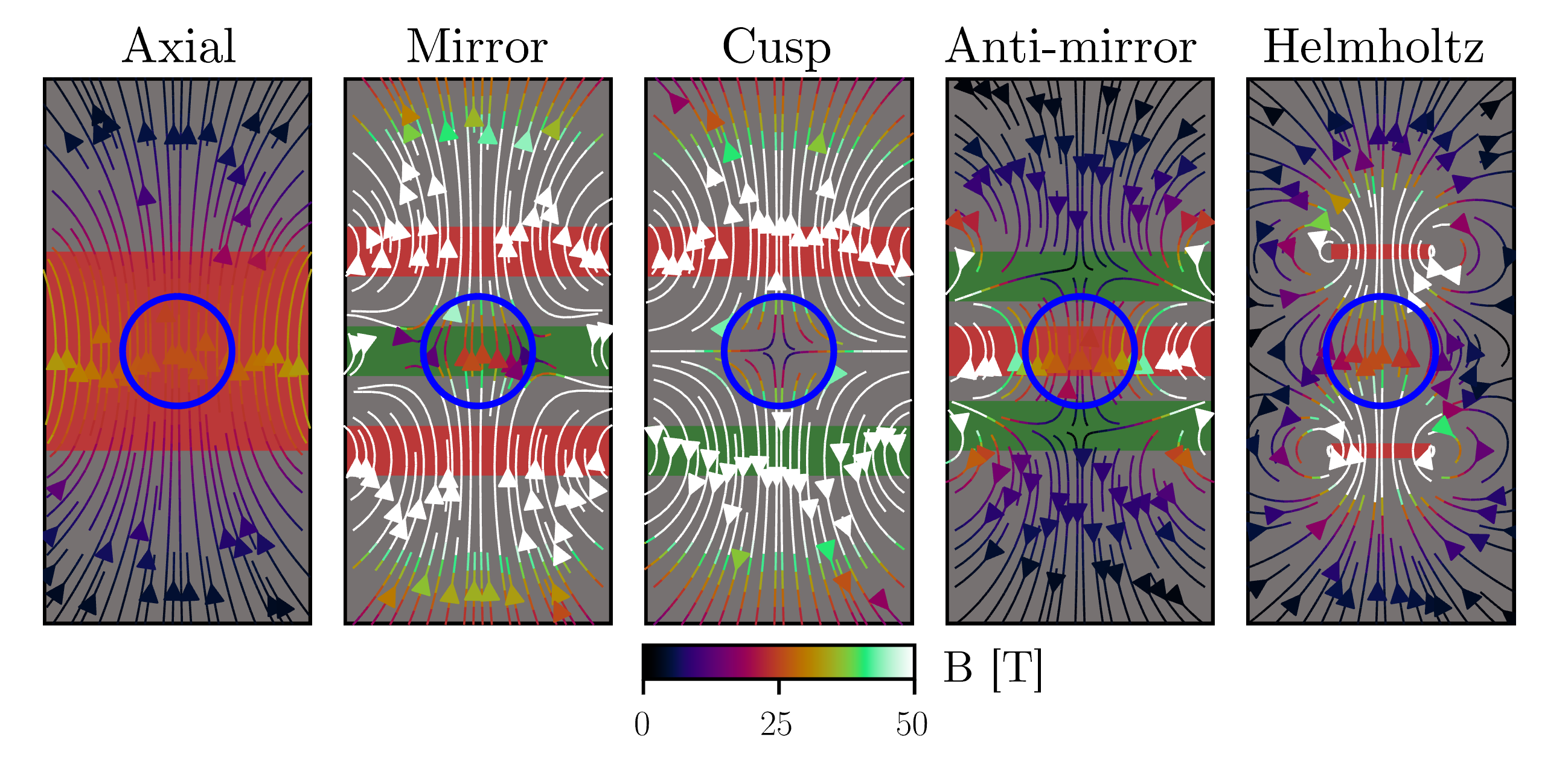} 
    \caption{Magnetic field profiles and coil diagrams for five field topologies explored in this section. Streamlines show the magnetic field direction and are color-coded to the field strength (in T). Fields are superimposed on a N210808-sized hohlraum (black rectangle) and initial capsule (blue circle). Red and green patches correspond to the solenoidal coil locations (radius, height, and position), while the color designates the current flow direction (red for counterclockwise, green for clockwise). For the mirror field, the center coil has 85\% the current of the outer coils, and the center coil of the anti-mirror has 200\% the current of its outer coils.} 
    \label{fig:coildiagrams} 
\end{figure*}

In this section, we demonstrate the resulting compressed field profiles of various non-axial field topologies using our advection-only model. One experimental consideration is that such fields must be produced using coils on the outside of a hohlraum (for indirect-drive) or with sufficient offset from a capsule to avoid beam clipping (for direct-drive). We consider the initial field topologies shown in Fig. \ref{fig:coildiagrams}, which are generated using a collection of finite-length solenoids \cite{callaghanMagneticFieldFinite1960}; the radius, length, and current direction are shown by the red and green patches on Fig. \ref{fig:coildiagrams}. The field profiles are demonstrated inside a N210808-sized hohlraum, showing how a mirror, cusp, and anti-mirror configuration might be produced. The current through each solenoid is tuned such that the volume-average field inside the capsule is 26 T, although high currents in the non-axial designs ($\sim 10-20$x higher than the axial design) may be infeasible. Furthermore, the interaction of each field topology with the hohlraum plasma to impact laser-plasma interaction and hot-electron preheat has not been studied except for the axial case \cite{strozziImposedMagneticField2015}. Finally, the ``Helmholtz'' design demonstrates how a mirror field might be applied to a direct-drive design, where the radius of the coils is not dictated by the hohlraum radius.

\begin{figure}[ht] 
    \centering 
    \includegraphics{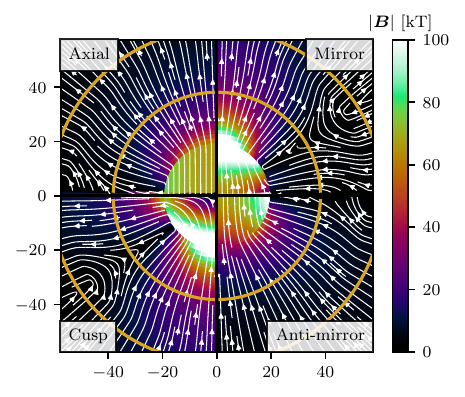} 
    \caption{Magnetic field profiles at peak compression for N210808-like implosions determined by applying the advection-only model (Eq. \ref{eq:evolutionBcyl}) to the initial non-axial field configurations in Fig. \ref{fig:coildiagrams}. Each quadrant corresponds to a different field configuration. Colored according to magnetic field strength (in kT).} 
    \label{fig:nonaxialheatmaps} 
\end{figure}

Using the advection-only model, the field profile of each of these topologies at peak compression is shown in Fig. \ref{fig:nonaxialheatmaps}. Irrespective of the initial field configuration, the field profile in the ablated ice is always predominantly radial and decays rapidly, suggesting that the initial field profile has negligible effect on the dynamics of the hotspot past the gas/ablated ice interface. However, in the zone of initial gas, the applied magnetic field retains its initial structure and may provide better thermal and alpha-particle insulation characteristics.

Thermal conduction in a magnetized plasma is anisotropic \cite{braginskiiTransportProcessesPlasma1965,epperleinPlasmaTransportCoefficients1986}, with cross-field transport becoming very small for these conditions with $\omega_{ce} \tau > 10$.
\begin{equation}
    \vv{q} = - \kappa_0 T^{5/2} \left( \left( \uv{b} \cdot \nabla T \right) \uv{b} + \frac{\kappa_\perp}{\kappa_\parallel}\uv{b} \times \left( \nabla T \times \uv{b} \right) \right)
\end{equation}
In this expression, $\uv{b}=\vv{B}/|\vv{B}|$ is the magnetic field unit vector, $\kappa_0 = (0.9842/\ln \Lambda) \ \mathrm{J / (ns \ \mu m \ keV^{7/2})}$ for $\bar{Z}=1$, and the thermal insulation factor $\kappa_\perp/\kappa_\parallel$ is given as a function of the Hall parameter $\omega_{ce} \tau$ by Ref. \onlinecite{epperleinPlasmaTransportCoefficients1986}. The outward thermal flux is determined by integrating $(\nabla \cdot \vv{q})$ over the spherical volume, giving 
\begin{multline} \label{eq:heatfluxaverage}
    \int \nabla \cdot \vv{q} dV = 
    - 2 \pi \kappa_0 T^{5/2} r^2 \times \\
    \int_0^\pi \left( \left( \uv{b} \cdot \nabla T \right) \uv{b} + \frac{\kappa_\perp}{\kappa_\parallel}\uv{b} \times \left( \nabla T \times \uv{b} \right) \right) \cdot \uv{r} \sin\theta d\theta \\
    \equiv - 4 \pi r^2 \langle \vv{q} \cdot \uv{r} \rangle.
\end{multline}
In the case of axial magnetic field ($\uv{b} = \uv{z}$) and radial temperature gradient ($\nabla T = \frac{\partial T}{\partial r} \uv{r}$), the average heat flux density $\langle \vv{q} \cdot \uv{r} \rangle$ has the well-known analytic solution \cite{walshMagnetizedICFImplosions2022,spiersHotspotModelInertial2025} of $\kappa_0 T^{5/2} \left( \frac{1}{3} + \frac{2}{3} \frac{\kappa_\perp}{\kappa_\parallel} \right) \nabla T$, implying that the magnetized total heat flux is just $1/3$ the equivalent unmagnetized heat flux.

\begin{figure}[ht] 
    \centering 
    \includegraphics{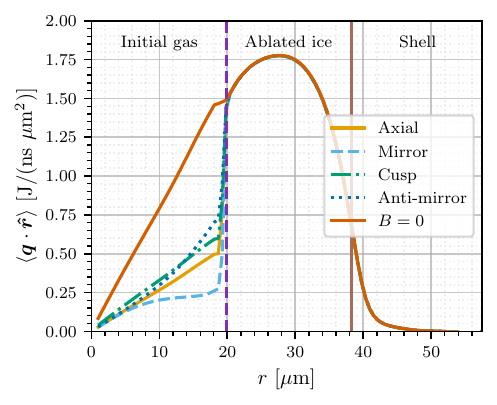} 
    \caption{Profiles of angular-averaged thermal conduction based on the field profiles in Fig. \ref{fig:nonaxialheatmaps}. Units are in terms of surface power density. Purple vertical line represents the interface between the initial gas and ablated ice, while the brown line is the hotspot/shell boundary.}
    \label{fig:nonaxialprofiles} 
\end{figure}

Thermal conduction profiles are demonstrated in Fig. \ref{fig:nonaxialprofiles} corresponding to evaluating expression \ref{eq:heatfluxaverage} for the field profiles given in Fig. \ref{fig:nonaxialheatmaps}, and assuming that $\nabla T$ is in the radial direction. Compared to the $B=0$ case, the thermal losses decrease by 60-80\% in the initial gas zone, depending on the field topology. Since the mirror field is more perpendicular to the radial direction, it shows the most thermal suppression, followed by the typical axial field. In the ablated ice region, the field is predominantly radial and therefore sees no thermal suppression due to the magnetic field.

\begin{figure}[ht] 
    \centering 
    \includegraphics{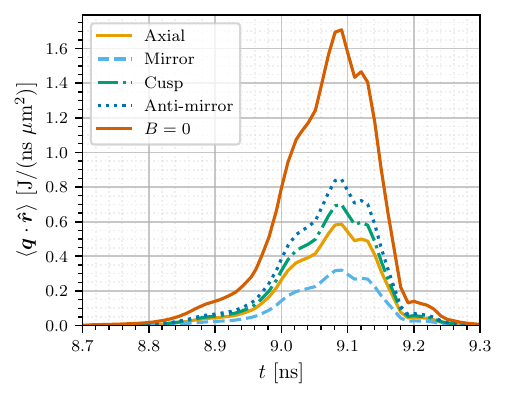} 
    \caption{Thermal conduction throughout the continuous deceleration phase near the interior of the gas/ablated ice interface for the advection-only solution to the magnetic field profiles.} 
    \label{fig:nonaxialhistory} 
\end{figure}

The time history of thermal conduction in the initial gas zone is shown in Fig. \ref{fig:nonaxialhistory}, demonstrating that thermal conduction is substantially suppressed throughout time for each of the magnetized configurations. This is because the electrons become magnetized as soon as the shock passes through the hotspot ($\sim 8.2$ ns) and have Hall parameters $\chi > 10$ throughout the continuous deceleration phase. Close to bangtime, compared to the unmagnetized case, thermal conduction is suppressed by 67\% for the axial field, 60\% for the cusp field, and 80\% for the mirror field. 

\section{Summary} \label{sec:summary}
This paper provides a mathematical solution for the 3D compression of an applied magnetic field in spherical ICF implosions, bridging the gap between oversimplified equatorial flux compression arguments and complex radiation-magnetohydrodynamics simulations. The model equations can be applied in general to any initial applied magnetic field that is compressed by a spherical implosion, thus serving as a robust tool for assessing the compressed field structure in ICF implosions.

We derive the evolution of magnetic fields in spherical flows (i.e., $\vv{u} = u_r (r, \theta, \varphi) \uv{r}$) which leads to a set of conservation laws (Eq. \ref{eq:evolutionBcyl}) that gives the leading-order magnetic field assuming that advection with the fluid dominates over feedback of the field on the implosion (e.g., via magnetic tension). We analyze characteristic implosion designs on the NIF and discuss the mechanisms causing fields to compress and bend, namely, we find that ablation from the shell into the hotspot induces a field-direction discontinuity at the initial gas/ablated ice interface. The model's flexibility and generality are demonstrated through application to asymmetric hotspots (mode-2, mode-4) and realistic non-axial applied magnetic fields, where we describe the feedback on hotspot thermodynamics via suppressed thermal conduction. We find that the magnetic mirror configuration may provide favorable insulation in the hotspot core but all cases are similar to unmagnetized in the ablated ice.

The next frontier of designing magnetized implosions is in finding an optimal target design over a vast parameter space. The analytic model in this work provides a rapid method for computing the 2D field topology for a given target and initial field design, enabling future high-dimensional parameter scans using only 1D simulations. Furthermore, the physics processes of magnetized implosions are intertwined (e.g., magnetization suppresses thermal conduction, which affects compression, thereby changing the field). In this work we decouple the magnetic field compression from the feedback on the implosion, providing a framework upon which to build a comprehensive theoretical understanding of magnetized hotspot physics. In subsequent work, we will model the physics of the rotational discontinuity to explain the complex hotspot field topologies exhibited in simulations of magnetized implosions. Then, we will present a simplified hotspot model which describes the coupled effects of field compression, magnetized thermal conduction and alpha-heating, and asymmetric hotspot evolution on the implosion yield, seeking an implosion design which maximizes the usefulness of the applied magnetic field.

\section{Appendix} \label{sec:appendix}
\subsection{Guidelines for using the field compression model} \label{subsec:usercode}

As part of this work, we have developed and published a simple python module (\texttt{field\_compression}) \cite{spiersBoseHEDPField_compression2026} which solves the equations of field compression in Sec. \ref{sec:theory}, allowing users to input arbitrary initial magnetic fields and density profiles, then outputs the evolved magnetic field according to Eq. \ref{eq:evolutionBcyl}. The model derived in this work has two primary assumptions: (1) only advection of the field with the fluid is considered, no feedback of the field on the hydrodynamics (e.g., $\vv{J} \times \vv{B}$) or thermodynamics (e.g., anisotropic thermal transport) is considered, and (2) the model is derived for quasi-spherical compression (i.e., $u_r(r, \theta, \varphi)$, see Eq. \ref{eq:conservationBr}), but simplifies under exact spherical flows (i.e., $u_r(r)$) as in Eqs. \ref{eq:evolutionBsph} and \ref{eq:evolutionBcyl}. The accompanying python module (\texttt{field\_compression}) has two modes of operation: trajectory evolution (requires $R_0$ and $R_f$) and density evolution (requires $\rho_0$ and $\rho_f$). These modes are mathematically equivalent, although density evolution is likely more convenient when the configurations originate from an Eulerian simulation or from analytic profiles.

\subsubsection{Trajectory evolution}

For this computation of the field compression, the only required variables are the relationship between the initial ($R_0$) and final ($R_f$) positions of each fluid element. To calculate the field bending parameter $\alpha$, one can either use the density profiles (like in Eq. \ref{eq:alpha}) or can numerically differentiate the fluid positions:
\begin{equation}
    \alpha \equiv 1 - \frac{\rho_f R_f^3}{\rho_0 R_0^3} = 1 - \frac{R_f}{R_0} \frac{\partial R_0}{\partial R_f}.
\end{equation}
The user then only needs to specify the initial magnetic field $\vv{B}_0 (\vv{r})$ in the desired coordinate system and the compressed magnetic field is computed at positions $R_f$ using Eq. \ref{eq:evolutionBcyl} (or, equivalently Eq. \ref{eq:evolutionBsph}).

\subsubsection{Density evolution}

Instead of tracking each fluid element throughout time, one can instead use only the density profiles $\rho_0$ and $\rho_f$ to determine the evolution of the fluid parcels. Using the Lagrangian continuity expression (Eq. \ref{eq:lagrangian}), this becomes a root-finding problem to determine $R_0$ given $\rho_0$, $\rho_f$ and any arbitrary $R_f$.
\begin{equation}
    \int_0^{R_f} \rho_f (x) x^2 dx - \int_0^{R_0} \rho_0 (x) x^2 dx = 0
\end{equation}
Now that $R_0$, $R_f$, $\rho_0$ and $\rho_f$ are known, specifying the initial magnetic field $\vv{B} (\vv{r})$ is sufficient to use Eq. \ref{eq:evolutionBcyl} to compute the compressed magnetic field at any location $R_f$.

\subsection{Vector potential derivation} \label{subsec:vecpot}

In Sec. \ref{sec:theory}, the solution to the advection-only induction equation under spherical flows ($\vv{u} = u_r(r, \theta, \varphi) \uv{r}$) was solved by expanding out the advection term and integrating by parts multiple times. Here, we exercise an alternative derivation which gives the same (albeit less general) result by using the vector potential. Such a solution may provide useful for validating or interpreting MHD codes that evolve the vector potential instead of the magnetic field components.

We also consider strictly \textit{poloidal} fields, i.e., fields in the $s-z$ (or $r-\theta$) plane, explicitly neglecting $B_\varphi$ in order to simplify the vector potential by gauge fixing. Furthermore, we assume that the implosion is spherically symmetric, i.e., $\vv{u} = u(r) \uv{r}$. Under these assumptions, we can rewrite the induction equation in terms of the azimuthal component of the vector potential $A_\varphi$ where $\vv{B} = \nabla \times \left( A_\varphi \uv{\varphi} \right)$. For example, an axial $\vv{B} = B_z \uv{z}$ has the corresponding vector potential $A_\varphi = \left( B_z / 2 \right) r \sin \theta$. The induction equation in terms of the vector potential is then
\begin{equation} \label{eq:inductionA}
    \frac{\partial \vv{A}}{\partial t} = \vv{u} \times \left( \nabla \times \vv{A} \right) = \left( \nabla \vv{A} \right) \cdot \vv{u} - \left( \vv{u} \cdot \nabla \right) \vv{A}.
\end{equation}

Using these constraints on $\vv{A}$ and $\vv{u}$, and making use of the convective derivative $\frac{D}{Dt} \equiv \frac{\partial}{\partial t} + \vv{u} \cdot \nabla$, the evolution of $\vv{A}$ for the spherically-convergent geometry is given by:
\begin{equation} \label{eq:inductionAconvective}
    \frac{D \vv{A}}{Dt} = -\frac{A_\varphi u}{r} \uv{\varphi}.
\end{equation}
Making use of the chain rule, it has been shown before by Chang \cite{changLaserDrivenMagneticFluxCompression2013} that this implies that $r A_\varphi$ is conserved with the flow.
\begin{equation} \label{eq:Aconservation}
    \frac{D }{Dt} \left( r A_\varphi \right) = 0
\end{equation}
The novel contribution by this derivation is to take the curl of $\vv{A}$ to determine the effect on the $\vv{B}$-field structure. Let us consider this problem in Lagrangian geometry, where the radial coordinates of the mesh move to conserve the differential mass enclosed. By differentiating our earlier definition of Lagrangian coordinates (Eq. \ref{eq:lagrangian}) with respect to $R_0$, we arrive at the typical expression for the Lagrangian Jacobian.
\begin{equation} \label{eq:jacobian}
    \frac{\partial R_f}{\partial R_0} = \frac{\rho_0 R_0^2}{\rho_f R_f^2}
\end{equation} 
Quantities are redefined in terms of Lagrangian coordinates and denoted with a tilde overhead, e.g. $A_\varphi (r, t)$ becomes $\tilde{A}_\varphi(R_0, t)$. Under this notation, integrating Eq. \ref{eq:Aconservation} over time yields 
\begin{equation} \label{eq:Alagrangian}
    \tilde{A}_{\varphi 0} R_0 = \tilde{A}_{\varphi f} R_f,
\end{equation}
indicating that the vector potential (and therefore $\vv{B}$) is path-independent, depending only on the initial and final Lagrangian cell positions. Taking the curl of $\vv{\tilde{A}}_0$ in spherical coordinates yields the initial $\vv{B}$-field in terms of spherical unit-vectors.
\begin{equation} \label{eq:curlA0}
\begin{split}
    \tilde{\vv{B}}_0 &= \nabla \times \tilde{\vv{A}}_0 \\
    &= \frac{1}{R_0 \sin \theta} \frac{\partial}{\partial \theta} \left( \sin \theta \tilde{A}_{\varphi 0} \right) \uv{r} - \frac{1}{R_0} \frac{\partial}{\partial R_0} \left( R_0 \tilde{A}_{\varphi 0} \right) \uv{\theta}
\end{split}
\end{equation}
The final magnetic field is determined by invoking the conservation of $A_\varphi R$ with the flow (Eq. \ref{eq:Alagrangian}) and taking the curl, where the resulting chain-rule is satisfied using Eq. \ref{eq:jacobian} to transform between $\partial/\partial R_f$ and $\partial/\partial R_0$.
\begin{equation} \label{eq:curlAf}
\begin{split}
    \tilde{\vv{B}}_f &= \nabla \times \left( \tilde{A}_{\varphi 0} \frac{R_0}{R_f} \uv{\varphi} \right) \\
    &= \frac{1}{R_f \sin \theta} \frac{\partial}{\partial \theta} \left( \sin \theta \tilde{A}_{\varphi 0} \frac{R_0}{R_f} \right) \uv{r} \\
    &\qquad - \frac{1}{R_f} \frac{\partial}{\partial R_f} \left( R_f \tilde{A}_{\varphi 0} \frac{R_0}{R_f} \right) \uv{\theta} \\
    &= \frac{R_0}{R_f^2 \sin\theta} \frac{\partial}{\partial \theta} \left( \sin \theta \tilde{A}_{\varphi 0} \right) \uv{r} - \frac{\rho_f R_f}{\rho_0 R_0^2} \frac{\partial}{\partial R_0} \left( R_0 \tilde{A}_{\varphi 0} \right) \uv{\theta}
\end{split}
\end{equation}
There is a clear correspondence between these equations for $\tilde{\vv{B}}_0$ and $\tilde{\vv{B}}_f$, represented by the matrix equation
\begin{equation} \label{eq:matrixBsph}
\begin{pmatrix}
    B_{r f} \\
    B_{\theta  f}
\end{pmatrix} = 
\begin{pmatrix}
    \left( \frac{R_0}{R_f} \right)^2 & 0 \\
    0 & \frac{\rho_f R_f}{\rho_0 R_0}
\end{pmatrix}
\begin{pmatrix}
    B_{r 0} \\
    B_{\theta 0}
\end{pmatrix},
\end{equation}
where $B_r$ and $B_\theta$ are the components of the magnetic field in the radial and polar-angle directions. Of more interest to magnetized-ICF is the magnetic field components in terms of the cylidrical unit vectors $\uv{s}$ and $\uv{z}$, which can be obtained by coordinate transformation. Upon transferring coordinates to cylindrical, we arrive back at Eq. \ref{eq:evolutionBcyl}.

\begin{acknowledgments}
    This material is supported by Department of Energy National Nuclear Security Administration Stewardship Science Graduate Fellowship program under award number DE-NA0003960, and through the University of Rochester Laboratory for Laser Energetics subaward number DENA0004144: SUB00000800/GR534362. This work was performed under the auspices of the U.S. Department of Energy by Lawrence Livermore National Laboratory under Contract No. DE-AC52-07NA27344 including LDRD project 23-ERD-025.
\end{acknowledgments}

\section*{Author Contribution}

\bibliography{fieldcompression}

\end{document}